\documentclass[11pt,letter]{article}
\usepackage[utf8]{inputenc}
\usepackage[]{algorithm2e}
\usepackage{physics}
\usepackage{amsmath}
\usepackage{amsfonts}
\usepackage{breqn}
\usepackage{graphicx}
\usepackage{xcolor}
\usepackage{hyperref}
\usepackage{color,soul}
 \usepackage{booktabs}
 \usepackage{url}
\usepackage[]{algorithm2e}
\usepackage{subfig}
\usepackage{authblk}

\usepackage{cite} 
\usepackage{authblk}
\usepackage{float}
\usepackage[export]{adjustbox}
\usepackage{rotating, graphicx}
\usepackage{lscape}  
\usepackage[left=2.5cm, right=2.5cm, top=2.5cm]{geometry}
\usepackage[export]{adjustbox}
\usepackage[final]{pdfpages}

\title{Quantum-resistance in blockchain networks}

\author[1,2]{M. Allende}
\author[1,2]{D. L\'opez Le\'on}
\author[1,2]{S. Cer\'on}
\author[1,2]{A. Leal}
\author[1,2]{A. Pareja}
\author[1,2]{M. Da Silva}
\author[1,2]{A. Pardo}
\author[3]{D. Jones}
\author[3]{D.J. Worrall}
\author[3]{B. Merriman}
\author[3]{J. Gilmore}
\author[3]{N. Kitchener}
\author[4]{S.E. Venegas-Andraca}

\affil[1]{IDB - Inter-American Development Bank, 1300 New York Ave, Washington DC, USA}
\affil[2]{LACChain - Global Alliance for the Development of the Blockchain Ecosystem in LAC}
\affil[3]{Cambridge Quantum Computing  - Cambridge, United Kingdom}
\affil[4]{Tecnologico de Monterrey, Escuela de Ingenieria y Ciencias. Monterrey, NL Mexico}

\begin{document}
\maketitle

{\small \tableofcontents}
\newpage{}

\begin{abstract}

This paper describes the work carried out by the Inter-American Development Bank, the IDB Lab, LACChain, Cambridge Quantum Computing (CQC), and Tecnologico de Monterrey to identify and eliminate quantum threats in blockchain networks. 

The advent of quantum computing threatens internet protocols and blockchain networks because they utilize non-quantum resistant cryptographic algorithms. When quantum computers become robust enough to run Shor's algorithm on a large scale, the most used asymmetric algorithms, utilized for digital signatures and message encryption, such as RSA,  (EC)DSA, and (EC)DH,  will be no longer secure. Quantum computers will be able to break them within a short period of time. Similarly, Grover's algorithm concedes a quadratic advantage for mining blocks in certain consensus protocols such as proof of work. 

Today, there are hundreds of billions of dollars denominated in cryptocurrencies that rely on blockchain ledgers as well as the thousands of blockchain-based applications storing value in blockchain networks. Cryptocurrencies and blockchain-based applications  require solutions that guarantee quantum resistance in order to preserve the integrity of data and assets in their public and immutable ledgers. We have designed and developed a layer-two solution to secure the exchange of information between blockchain nodes over the internet and introduced a second signature in transactions using post-quantum keys. Our versatile solution can be applied to any blockchain network. In our implementation, quantum entropy was provided via the IronBridge Platform from CQC and we used LACChain Besu as the blockchain network.

\end{abstract}

\section{Introduction}

Quantum computing, one of the most recent cross-pollination efforts between physics and computer science, is a scientific and engineering field focused on developing information processing devices and algorithms based on quantum mechanics \cite{benioff80,benioff82a,feynman82,feynman86,deutsch00,feynman_lectures_computation,deutsch85}. Quantum computing is now an established research field with solid theoretical and experimental results \cite{kadowaki98,aharonov07,mcgeoch-qa,venegas-andraca-qw-12,biamonte21}. Furthermore, high-tech businesses across various sectors are increasingly experimenting with quantum computing technological solutions \cite{cambridgeqc,multiverse,alex21,winiarczyk13}.

Since the early days of quantum computing, the role of quantum algorithms and quantum protocols in information security has been a crucial issue. On the one hand, Shor's algorithm \cite{shor97} could be used to break public-key cryptography protocols. On the other hand, Quantum Key Distribution schemes provide security levels to information transmission that are not based on mathematical conjectures but instead on the properties of quantum mechanics \cite{loepp2006}. Quantum entropy provides perfect randomness and strong cryptographic keys based on quantum mechanics \cite{cqc-entropy}.  Post-Quantum Cryptography encompasses a new generation of algorithms for the creation of asymmetric keys that are thought to be resistant to attacks by quantum computers \cite{bernstein2017}.

Currently, blockchain \cite{haber91} is the most popular technology amongst emerging applications for decentralized data sharing and storage. The design and implementation of blockchain networks makes extensive use of cryptography protocols; thus, studying the potential uses of quantum computing to both weaken and strengthen blockchain technologies is essential to ensuring its future reliability.

\newpage{}

The rest of this paper is divided as follows. Section \ref{context} presents an introductory review of Quantum Computing, Quantum Key Distribution, Post-Quantum Cryptography, blockchain, and the LACChain Blockchain Network; Section \ref{vulnerabilities} analyzes relevant vulnerabilities of blockchain within the context of quantum computing technologies; Section \ref{proposal-quantum-safe-blockchain} introduces our solution for guaranteeing quantum-resistance in blockchain networks and describes the implementation carried out in the LACChain Blockchain Network; Section \ref{proposalimplementation} explores several key implementation matters; Section  \ref{conclusions} presents conclusions and future directions.

\section{Context}
\label{context}
\subsection{Quantum computing as a threat to cryptography}

Theoretical results, such as Shor's algorithm \cite{shor97}, and state-of-the-art quantum computing technology in conjunction with expected near-to-mid future scalability and robust developments  have attracted the attention of international standards agencies in cyber security and cryptography, including NIST \cite{nist-report}, NSA\cite{nsa-report}, and ETSI\cite{etsiwp8}. They have made critical warnings that running some quantum algorithms on full-scale quantum computers will necessitate the protection of internet and telecommunication information exchanges for widely used cryptography protocols. Most notably, NIST is currently  running a post-quantum cryptography competition for standardization to replace existing cryptographic algorithms that are susceptible to breakage using quantum computers \cite{nist-pqc}.

Quantum computers use quantum bits (qubits) as fundamental units of information. Individual qubits can be in binary zero and one states (classical bits), but they can also be in any state between zero  and one, which is defined by  the superposition $\alpha |0\rangle + \beta |1\rangle$ where $\alpha, \beta \in \mathbb{C}$ subject to $|\alpha|^2 + |\beta|^2 =1$. Qubits leverage the quantum effects that do not appear in classical computing, such as quantum superposition, quantum entanglement and quantum tunneling. These effects are fundamental for the development of quantum algorithms, which have proven to be very useful in solving certain problems much more efficiently than the best-known classical algorithms, such as optimization or factorization of prime numbers.

In general,  physical channels currently used to transmit digital information are unprotected (e.g., optical fibers or wireless transmissions) and the security of data exchanges within these channels relies on cryptographic protocols. 
It is only a matter of time before large and robust quantum computers capable of breaking current cryptographic protocols are built. It is crucial that we be prepared for these future technologies, especially in order to investigate the transition to quantum-safe cryptography for blockchain technologies.

\subsection{Current approaches for quantum-safe cryptography}
Discussions on quantum computers and cryptography usually surround two main areas of cryptography that are thought to resist attacks by large and robust quantum computers: quantum key distribution and post-quantum cryptography. 

\newpage{}

\subsubsection{Quantum Key Distribution}
\label{subsection-qkd}

Quantum Key Distribution (QKD) refers to quantum protocols for the co-creation of private symmetric keys between two parties using quantum and classical channels (e.g., optical fibers and wireless channels) by codifying private key bits into quantum states. If these quantum states are intercepted and observed by any eavesdropper, the information they contain (i.e., the bits of the key) is modified, and therefore the key is corrupted and the eavesdropper is detected. Best known QKD protocols are BB84 \cite{bb84,bb84thirtyyearslater} and E91 \cite{ekert91}.

An illustrative example of a QKD implementation is the BB84 protocol using polarized photons. In this protocol, we have a sender (Alice), a recipient (Bob), and an eavesdropper (Eve). Alice codes the bits of a private key to share with Bob using non-orthogonal quantum states, such as bit value $0$ using either $|0 \rangle$ or $|+ \rangle$ and bit value $1$ using $|1 \rangle$ or $|- \rangle$. Then, photons are sent by Alice to Bob. Due to the properties of measurement in quantum mechanics, Eve's eavesdropping activities will eventually be detected (that is, Eve's activities will  leave a trace that will eventually be detected by Alice and Bob) and, consequently, the protocol will stop and start over at a later stage \cite{bouwmeester01,bassem2018}.

QKD protocols such as BB84  and E91  have been successfully implemented since 2003. However, QKD is not fully scalable today because ground-based key exchanges using optical fibers are limited to a few hundreds kilometers due to the degradation of the quantum states containing the keys \cite{lucamarini18}. Additionally, ground-to-satellite key exchanges require sophisticated infrastructure for generation, transmission, and reception of quantum keys \cite{liao17,li19}. The scalability of these networks depends on the development of quantum repeaters, which require very sophisticated quantum memories. This is still an area under development \cite{focusquantummemories,heshami16}. For these reasons, QKD has been discarded as a feasible solution to provide quantum safeness to blockchain networks today.   However, this may change in the future as NSA, NIST, and ETSI, among others, have declared that quantum cryptography (such as QKD) would be the only alternative for long term secure encryption \cite{nsa-report, nist-report, etsiwp8}.  

\subsubsection{Post-Quantum Cryptography}
\label{subsection-pqc}

Existing symmetric standards such as AES have already well-understood variants that are believed to provide adequate security against quantum adversaries. In contrast, it is well known that public (asymmetric) key cryptographic protocols such as RSA \cite{Rivest78amethod,rsapatent}, (Elliptic Curve) Digital Signature Algorithm \cite{smart2016}, and (Elliptic Curve) Diffie-Hellman \cite{dh76,miller1986} are considered vulnerable to quantum attacks.

Post-Quantum Cryptography (PQC) refers to a new generation of asymmetric algorithms that cannot be broken by Shor's algorithm. Unlike QKD, PQC does not rely on any underlying quantum processes but rather on more complex mathematical problems. The main focus areas for post-quantum algorithms to generate quantum-safe asymmetric key pairs are:

\begin{itemize}
\item
Hash-based Cryptography, based on the security of hash functions.
\item 
Code-based Cryptography, based on the difficulty of decoding generic linear code.
\item
Lattice-based Cryptography, based on the difficulty of well-studied lattice problems (e.g., shortest vector problem).
\item
Multivariate Cryptography, based on multivariate polynomials over a finite field.
\end{itemize}

\newpage{}

As mentioned above, there is a standardization process being conducted by NIST which started in August 2016 with a request for comments \cite{nist-pqc}. This process, which called for submissions in the areas of \lq \lq Public-key Encryption and Key Establishment Mechanisms (KEM)'' and \lq \lq Digital Signature Algorithms'' announced the final and alternate rounds of in July 2020 \cite{nist-round3-submissions}. The final algorithms are estimated to be standardized between 2022 and 2024 \cite{nist-timeline}.  There are various initiatives running alongside NIST's initiative such as PQCrypto \cite{pqcrypto} and Open Quantum Safe \cite{oqs}. NITS's finalists in the KEM category are:

\begin{itemize}
\item
Classic McEliece, a code-based scheme.\cite{mcelieve}.
\item
Crystals-Kyber, a suite of algebraic lattices utilizing a Kyber primitive for KEM  \cite{crystalskyber}.
\item
NTRU, a lattice-based scheme  \cite{ntru}.
\item
Saber, a lattice-based scheme utilizing learning with rounding  \cite{saber}.
\end{itemize}

 The Digital Signature Algorithms are: 
 \begin{itemize}
\item
Crystals-Dilithium, a suite of Algebraic lattices using a Dilithium primitive for signature  \cite{crystalsdilithium}.
\item
Falcon, lattice-based algorithm with shake256 hashing  \cite{falcon}.
\item
Rainbow, multivariate based solution  \cite{rainbow}.
\end{itemize}

There are also a number of alternates proposed for both categories. Comments on the submissions' security and efficacy can be found in \cite{commentsnist}. While there are several candidates sharing a similar approach, their proposals vary in key sizes and signature sizes, making it necessary to evaluate each scheme against the architecture in which candidates are intended to be deployed.

\subsection{Blockchain and the LACChain Blockchain Network}

Blockchain is a technology that allows one to build decentralized ledgers in which different entities can register transactions that are grouped into blocks that are linked using hashes \cite{haber91}. The immutability of the transactions stored in blockchain networks is guaranteed because it is impossible to tamper with the ledger without being detected. As any entity can, in principle, have a synchronized copy of the ledger and transactions that are validated according to predefined rules, the history cannot be rewritten. The integrity of the transactions is guaranteed by digital signatures because  every transaction is signed by the sender, and the immutability of the chain is guaranteed by hash functions \cite{haber91}.

Our work analyzes vulnerabilities of hash functions and cryptographic algorithms. The security of these core elements of blockchain networks will be threatened when quantum computers become robust enough. This applies to most blockchain networks and it is a critical concern that the blockchain community has not yet properly addressed.

Vitalik Buterin, one of the founders of the Ethereum blockchain technology, acknowledged the quantum threat back in 2015 and suggested moving towards Lamport signatures eventually \cite{vitalik}. Prior to our work, the University of Waterloo and Microsoft Research estimated that the number of logical qubits necessary to implement quantum algorithms that can break 256 bit-long digital signatures generated with  (EC)DSA, typically used in current blockchain networks, are 1500 \cite{proos} and 2330 \cite{roettleler}, respectively. It is still unclear how many physical qubits would be needed for that purpose. Another study by researchers in Singapore, Australia, and France claimed in 2017 that quantum computers will be large and robust enough to break Bitcoin keys in 10 minutes by 2017 \cite{aggarwal}. In 2018, three groups of scientists from Russia and Canada achieved an  implementation of a quantum-secured blockchain based on an exchange of keys using QKD techniques \cite{kitkenko}, but their scalability is limited by the limitations of the channels for QKD exchange. Additional work has been published since these initial analyses \cite{birmingham,shen,meryem,alexandru,jiahui,wei}. However, we are not aware of any scalable implementation of a quantum-safe blockchain network prior to our work.

We have designed a solution that can be deployed in different blockchain networks. As a key component to show the viability of our proposal, we have implemented it in the LACChain Consensys Quorum (a.k.a. Besu) Network. LACChain is a blockchain infrastructure led by the Innovation Lab of the Inter-American Development Bank (IDB Lab) in Global Alliance with some of the entities leading the development of blockchain technology in the world \cite{lacchain21}. The main goal of LACChain is to enable a robust and scalable blockchain network that can host multipurpose use cases with social, economic, and financial impact. Hyperledger Besu is an Ethereum client originally developed by Consensys and now maintained by the Ethereum community, including Consensys \cite{besu}. 

Blockchain can be thought of as a computational system with a distributed state shared among a network of nodes, of which consistency can be verified by any participant. The state is dynamically updated through messages, called transactions, that are broadcasted by the nodes, and each participant can have a verified and verifiable copy of the state and the transaction history. These transactions allow users to deploy executable code to the network, a.k.a. smart contracts, and interact with them. 

In order for a new state to be agreed upon by the network, a subset of nodes, called validator or producer nodes, apply a consensus protocol. There are different types of consensus protocols and each network decides which type of consensus protocol they implement. Essentially, every consensus protocol consists of a set of rules that establish how these nodes will accomplish a computational validation of the latest transactions replicated across the network. The validator or producer nodes propose a package, called a block, which contains the transaction, block number, nonce, block hash, previous block hash,  and signatures of the block validators or producers. With this, a new block is cryptographically sealed and, once appended to the blockchain, it cannot be undone or tampered with.

In Ethereum Networks, the code deployed in the network is a stream of bytes representing operation codes from the Ethereum Virtual Machine (a.k.a. EVM). This set of operations can be considered Turing complete and are executed as a stack machine with a depth of 1024 items. The EVM is then the runtime environment where any state transformation takes place \cite{ethereumevm}. Every smart contract has its own memory space and can be changed or updated by a transaction, which is recorded in the transaction history and implies a modification of the current distributed state. Additionally, each operation has an associated cost, which is an abstraction of the computational power required to perform the requested action by an ideal computer. The cost is called gas and serves as a metric for the amount of computation required to process each block.

\newpage{}

\section{The vulnerabilities of blockchain technology with the advent of quantum computing}
\label{vulnerabilities}

The advent of quantum computing constitutes a new paradigm in which digital technologies will endure both challenges and opportunities. Threats will come up in a variety of forms, especially when robust quantum computers will be able to break several important cryptographic algorithms currently used. Blockchain, as a technology that strongly relies on cryptography, is not safe from these threats. As stated in \cite{allende19}, it is worth exploring the conjunction of blockchain technology and quantum computing in the following  four areas.

\begin{itemize}

\item 
Digital signatures are one of the most essential components of blockchain technology.  Bitcoin and Ethereum use elliptic curve cryptography (ECC), particularly the ECDSA signature schemes on curve secp256k1. Others, such as EOSIO, use the NIST standard secp256r1 curve.  NIST recommends that ECDSA and RSA signature schemes be replaced due to the impact of Shor's algorithm on these schemes \cite{nistreportpqc16}.

\item 
Communication over the Internet relies on protocols such as  HTTP. The security of the communication happens in HTTPS within the SSL/TLS protocol stack. TLS supports one-time key generation (which is not quantum safe) with AES for symmetric encryption and several non-quantum-safe algorithms for exchange and authentication, such as RSA, DH, ECDH, ECDSA, and DSA. This means that all internet communications, including transactions and messages sent between applications and nodes in a blockchain, will not be quantum safe when robust quantum computers become fully operational.


\item
Block mining: blockchain networks that use proof-of-work as the consensus mechanism rely on finding nonces. Quantum computers will be able to find these nonces quadratically faster using Grover's algorithm \cite{grover96}. However, this does  not  pose  a  major  threat  to  the  security of blockchain networks because the solution will be as easy as quadratically increasing the difficulty to compensate for the quantum advantage. In networks with consensus protocols that do not promote competition between nodes, such as the proof-of-authority used in the LACChain Blockchain, this threat will not exist.

\item

Hash functions take an element from a set of infinitely many elements and gives an output from a finite set of $2^{256}$ elements in the case of the SHA-256 function that is used by most of the blockchain networks today. Thus, from a hash value stored in the blockchain, it is statistically impossible to obtain the element that resulted in that value. This property, known as irreversibility or pre-image resistance, guarantees the security of these operations even in the presence of quantum computers \cite{allende19}. 

Additionally, hash functions are continually evolving for increased security. For example, if quantum computers evolve to the point of posing a threat to SHA-2, then SHA-3 is already standardized as an alternative that offers a higher level of security in NIST standard FIPS202 \cite{nistfips202}. 

\end{itemize}

\newpage{}

\section{A Proposal for a Quantum-Safe Blockchain Network}
\label{proposal-quantum-safe-blockchain}
As a result of this high-level analysis, it becomes clear that the threat blockchain networks face with respect to quantum computers is primarily related to vulnerable digital signatures of blockchain transactions and vulnerable key-exchange mechanisms  used for the peer-to-peer communication over the network.  The solution we propose does not require modification of the algorithms used by the Internet or blockchain protocols but creates a layer on top that provides quantum security. This solution consists of:

\begin{itemize}

\item
Encapsulating communication between nodes using post-quantum X.509 certificates  to establish TLS tunnels. As part of the on-boarding process, nodes are issued a \lq \lq post-quantum X.509 certificate'', from a LACChain Certificate Authority (CA), which is an extension of an X.509 certificate using the v3 extension specification  that allows for the incorporation of new fields into the credential, such as complementary cryptographic algorithms. In our case,  these complementary algorithms are post-quantum \cite{x509v3}. Using these certificates, nodes can establish secure post-quantum connections that encapsulate data sharing over the communication protocol, defined by the blockchain network. The encapsulated data are transactions broadcasted by writer nodes and the blocks produced by producer or validator nodes.

\item
Signing transactions with a post-quantum signature along with the regular signature defined in the blockchain protocol and establishing on-chain verification mechanisms. Our solution consists of enabling a second layer cryptography scheme that allows nodes that broadcast transactions -writer nodes- to sign them with a post-quantum signature that can be verified on-chain. This is in addition to the ECDSA signature that comes by default with the blockchain protocol. If the ECDSA signature becomes compromised by a quantum computer, integrity is preserved by the post-quantum signature. We leverage the post-quantum keys associated to the post-quantum X.509 certificates for this purpose. 
\end{itemize}

For both the encapsulation and transaction signing, we rely on certified quantum entropy to generate keys for maximal security.

One could argue that by the time large quantum computers capable of breaking current cryptography are ready, blockchain protocols will have upgraded their cryptography to post-quantum safe algorithms. However, considering that blockchain networks are immutable ledgers, the rule of \lq \lq hack today, crack tomorrow'' urges us to protect them now.

For example, a university can start issuing digital diplomas today and register the proofs with their digital signature (ECC or RSA) in a blockchain network. However, in 5, 10, or 15 years, when a quantum computer can break that signature and discover the private key, all previously issued digital diplomas will be compromised, as the issuer can be impersonated. Further, we there is no way of knowing whether a person has a quantum computer with the capacity to impersonate others and steal their assets without being detected. The same rationale can be applied to the issuance of a bond or the issuance of a central bank digital currency (CBDC) by a Central Bank.

\newpage{}

\section{Implementation}
\label{proposalimplementation}
The implementation of our solution is composed of the following five phases:

\begin{enumerate}

\item
Generation and distribution of quantum entropy.

\item
Generation of post-quantum certificates.

\item
Encapsulation of the communication between nodes using quantum-safe cryptography.

\item
Signature of transactions using post-quantum keys.

\item
On-chain verification of post-quantum signatures.

\end{enumerate}

\subsection{Generation and distribution of quantum entropy}
\label{gen-dist-quantum-entropy}

Randomness is the cornerstone upon which cryptographic standards are built. It is used to generate the keys and seeds used in cryptographic schemes. The challenge related to the generation of randomness is the generation of truly random data. Current techniques rely on deterministic approaches - hardware utilizing classical physics, and any available inputs that might add some level of unpredictability - which leads to the generation of pseudo-random data in the vast majority of the cases. Failure to ensure sufficient randomness in cryptographic processes can lead to real-world attacks on otherwise secure systems. This even extends to quantum random number generators which is why there is a need to develop schemes for true randomness \cite{zheng2020}.

Conversely, quantum generation of randomness harnesses the power of the non-deterministic nature of quantum mechanics. Generating quantum random numbers \cite{herrero2016}  can be built in many ways, as has been illustrated by the various approaches used to date, including beam splitters with detectors, vacuum fluctuations in coherent light, and squeezed coherent light mechanisms, among others \cite{shi16,leone20}. Despite the fact that these methods are non-deterministic, they lack the ability for an end user to guarantee that the device is working correctly. This ability in a device (sometimes known as device independence or more commonly, as certifiably quantum generation) is at the heart of the qRNG, Ironbridge, used in our solution presented in this paper.

IronBridge generates randomness through a quantum process evaluated as quantum certifiable which utilizes a test for the violation of a Bell Inequality \cite{bell87,bellstanford} or a higher order test of a Mermin Inequality on a NISQ machine \cite{huang20}. Such a violation, along with various other security tests, are taken as mathematical proof that the output could have only come from a quantum source and is non-deterministic and thus maximally random for a physical system. For the experiments in this paper, an IBM quantum computer was used to generate the entropy.

Given the distributed nature of a blockchain, ideally each entity running a node should have its own local source of quantum entropy: a qRNG device. However, it was not feasible to provide each node with its own qRNG for our pilot, so we used a central source of quantum entropy. As discussed throughout this paper, current cryptographic schemes used in SSL/TLS are not quantum-safe, so using them to distribute the entropy would have broken the quantum-safeness at the start. 

We decided instead to design a protocol that allowed nodes to create a quantum safe tunnel between themselves and the entropy distribution point  to ensure that this communication could be considered quantum safe. In order to do this, the entropy source creates a first key, splits it into several parts, and delivers it to the node through various TLS channels. Nodes have a time out to receive the key, recompose it, and use it to authenticate against the entropy source.  This is covered in more detail in Section \ref{entsourcesetup}.

\subsubsection{OpenSSL Framework}

Over the last 20 years, the OpenSSL API has become the de-facto cryptographic framework for applications that use TLS/SSL, providing capabilities such as:

\begin{itemize}
\item 
Generation of pseudo-random numbers.
\item
Classical cryptographic support using algorithms such as Diffie-Hellman (DH) and Elliptic Curve Diffie-Hellman (ECDH).
\end{itemize}

The OpenSSL applications and libraries also provide the following functions:

\begin{itemize}

\item 
Generation of private and public key pairs.

\item 
Certificate authority management.

\item 
Certificate validation.

\item 
Management of crypto libraries and engine plugins to support new algorithms.

\item
SSL/TLS client and server implementations.

\end{itemize}

Because quantum computing will impact the security of asymmetric cryptographic algorithms such as RSA and ECDSA, the following changes within OpenSSL are required:

\begin{itemize}

\item 
Support for certified quantum entropy to replace the existing pseudo-random number generator used to seed keys and random values used for nonce parameters.

\item 
Support for post-quantum algorithms to provide both key encapsulation and digital signatures.

\end{itemize}

IronBridge Platform facilitates the move to OpenSSL with entropy provided for:

\begin{itemize}

\item 
Quantum key encapsulation protecting existing PKI infrastructure by wrapping non-post quantum resistant keys in a post quantum wrapper.

\item 
Quantum generated random numbers for pure quantum generated keys for signature digest algorithms.

\end{itemize}

This approach facilitates easy integration into computer security layers within the operating system while still being compatible with most of the existing infrastructure. The IronBridge Service Agent provides post quantum encapsulated key management for the secure entropy tunnel back to the IronBridge Platform. The component provides users with the ability to enforce customer security policies with regard to maximum key lifetimes by automatically providing configurable key cycling capability.

\subsubsection{Entropy Source Setup}
\label{entsourcesetup}

As mentioned before, every blockchain node should ideally have its own source of quantum entropy. For our pilot, LACChain nodes did not have a local source of quantum entropy so it was necessary to establish a quantum-safe connection between the external source (the IronBridge platform) and each of the nodes. As the quantum entropy is necessary to generate the post-quantum keys that allow establishment of a quantum-safe connection, we could not use post-quantum cryptography to protect this first channel.

Therefore, we designed a protocol that begins with the distribution of a post-quantum key from the IronBridge platform to the LACChain nodes. This key is split into N parts and delivered through different TLS channels. Once the LACChain node is in possession of all the N parts, it reconstructs the key and uses it to establish a first connection with the quantum entropy source. This key is only used once, and afterwards it is immediately discarded.

CQC IronBridge provides certified quantum generated entropy for cryptographic use, delivering stronger classical cryptography and the highest strength post-quantum cryptography within customer's cryptographic ecosystems.  IronBridge's patent-pending device independent certification mathematically proves every random number is the outcome of a quantum process without trusting the generation process before customer use.

\begin{figure}[hbt]
\includegraphics[width=1\textwidth,left]{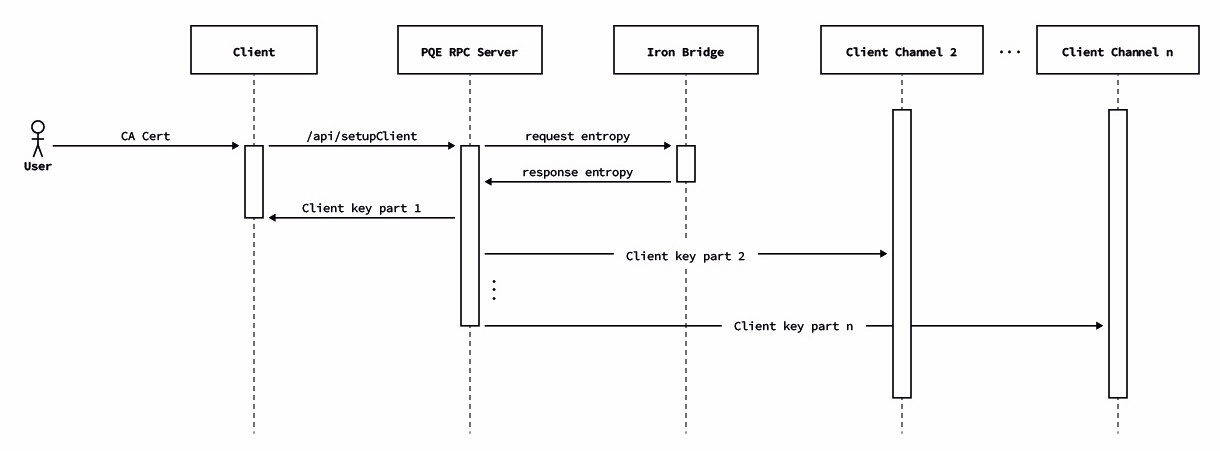}
\caption{{\small High-level schema of the first connection between the remote source of entropy and the blockchain node.}}
\label{figure-01}
\end{figure}

Once this first post-quantum key is used to establish the first secure connection between the LACChain node and the entropy source, they initiate a second process to renegotiate a working KEM keypair using the post-quantum algorithm, McEliece, in line with the NIST round three submissions \cite{nist-round3-submissions}. This allows for the establishment of a quantum-safe connection between the entropy source and the nodes which allows the LACChain nodes to start requesting quantum entropy on demand (see Fig. (\ref{figure-01})).


\newpage{}
\subsection{Generation of Post-Quantum Certificates}
\label{generationofpqcs}





\begin{figure}[hbtp]
\begin{center}
\resizebox{135mm}{!}{\includegraphics{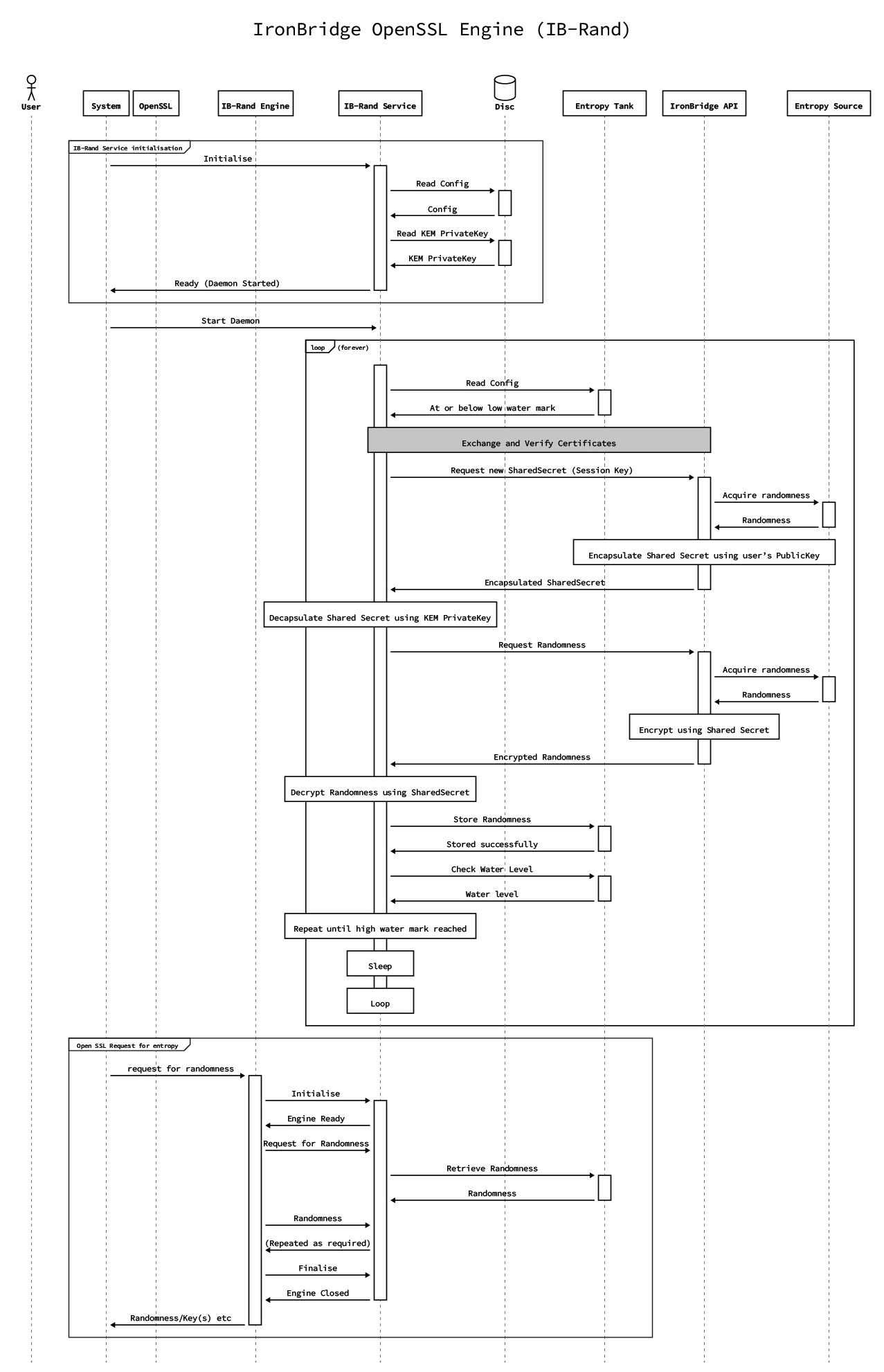}}
\end{center}
\caption{{\small Detailed flows describing the generation and consumption of entropy on demand by the Open SSL.}}
\label{figure-02}
\end{figure}

Once the LACChain nodes have access to quantum entropy on demand, this entropy is consumed by OpenSSL as illustrated in Fig. (\ref{figure-02}). Permanent quantum-safe cryptographic solutions such as QKD (see Section \ref{subsection-qkd}) are not scalable today and require substantial investments in infrastructure. Feasible and practical solutions that provide quantum-resistance today involve PQC (see Section \ref{subsection-pqc}). Instead of replacing current Internet and blockchain protocols with new ones that incorporate PQC, we tried to introduce PQC in existing frameworks.


Based on the analysis presented above, we decided to use the traditional X.509 standard, which defines an internationally accepted format for digital documents that securely associates cryptographic key pairs with identities such as websites, individuals, and organizations \cite{rfc5280website}.

By using a modified version of libSSL, the X.509 specification was extended to incorporate post-quantum and Ethereum (ECDSA) public keys, allowing blockchain nodes to use the modified libSSL to establish peer-to-peer quantum-safe channels that leverage those keys. Libssl is the portion of OpenSSL  that supports TLS (SSL and TLS Protocols) and depends on libcrypto.

As discussed in Section \ref{proposal-quantum-safe-blockchain},  the nodes use the post-quantum keys to encapsulate communication with other nodes and sign transactions broadcasted to the blockchain. We decided to use the same algorithm for the generation of both types of keys (i.e., encryption keys and signing keys). Given the versatility of OpenSSL to incorporate any post-quantum algorithm, the election of the post-quantum algorithm was based on the restrictions inherent in executing blockchain transactions -essentially execution time and payload size- as different algorithms present substantial differences that condition the feasibility of on-chain verifications and storage.

\begin{figure}[hbtp]
\includegraphics[width=0.8\textwidth,center]{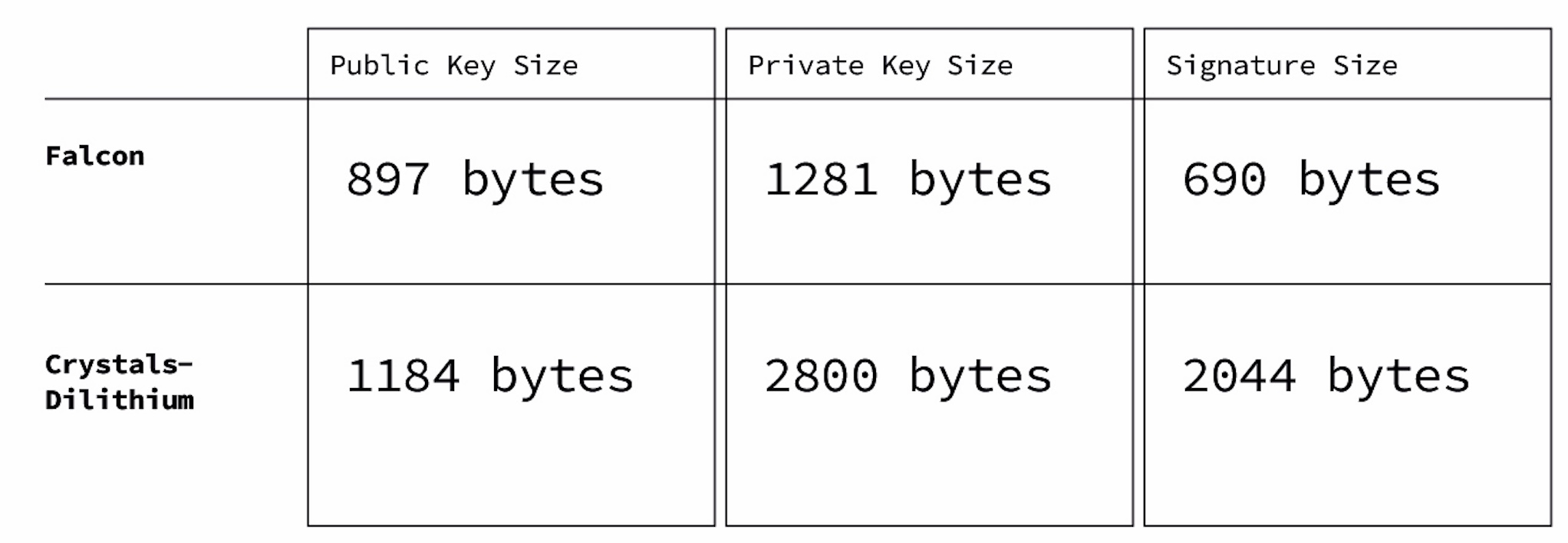}
\caption{{\small Comparison between Falcon and Crystals-Filithium algorithms.}}
\label{figure-03}
\end{figure}

We evaluated the two finalists of the NIST competition in the signature category \cite{nist-round3-submissions}, Crystals-Dilithium \cite{crystalsdilithium} and Falcon \cite{falcon}. Figure \ref{figure-03} presents some of the differences between these two algorithms in terms of public key size, private key size, and signature size.

Both algorithms are very demanding regarding processing, memory, and amount of random material required to compute keys and signatures. However, Falcon has been acknowledged as the most compact and contains a built-in SHA3 compliant Extendable Output Function (XOF Shake256). The Ethereum VM natively supports the Keccak hashing algorithm upon which SHA 3 NIST FIPS202 is based, but it does not provide the extendable output functions (XOF) required. Further, implementing the shake XOF functionality is not straightforward.

We evaluated the other signing algorithms but speed, complexity,  and the  fact that we would have to implement a SHA3 compliant ecosystem for the qRNG source to feed those schemes proved Falcon to be the best option. Our solution allows for the incorporation of new post-quantum algorithms, such as those that can be standardized by organizations such as NIST in the upcoming months and years.

To use Falcon, we needed to add a new object identifier (OID), the 1.3.9999.3.1, to libSSL in order to recognize the post quantum Falcon-512 algorithm \cite{falcon-github}.

\begin{figure}[hbt]
\includegraphics[width=1\textwidth,center]{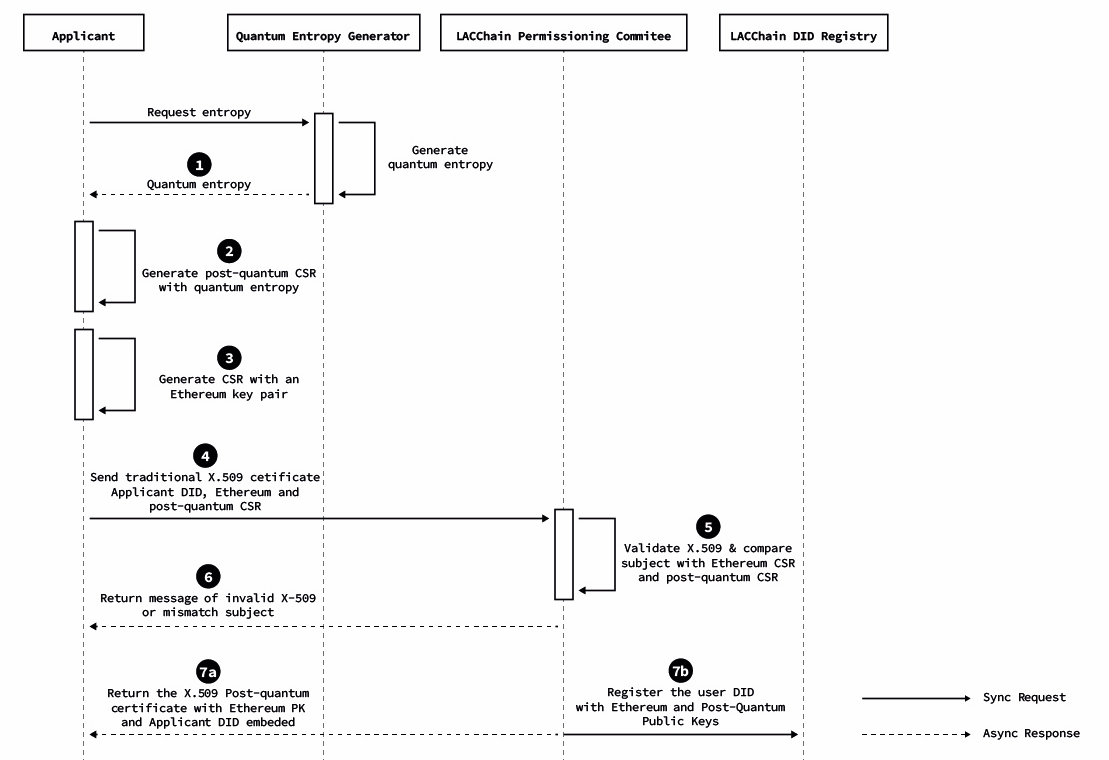}
\caption{{\small High level diagram of the post-quantum certification and on-chain registration of an entity.}}
\label{figure-04}
\end{figure}

\begin{figure}[hbt]
\includegraphics[width=1\textwidth,center]{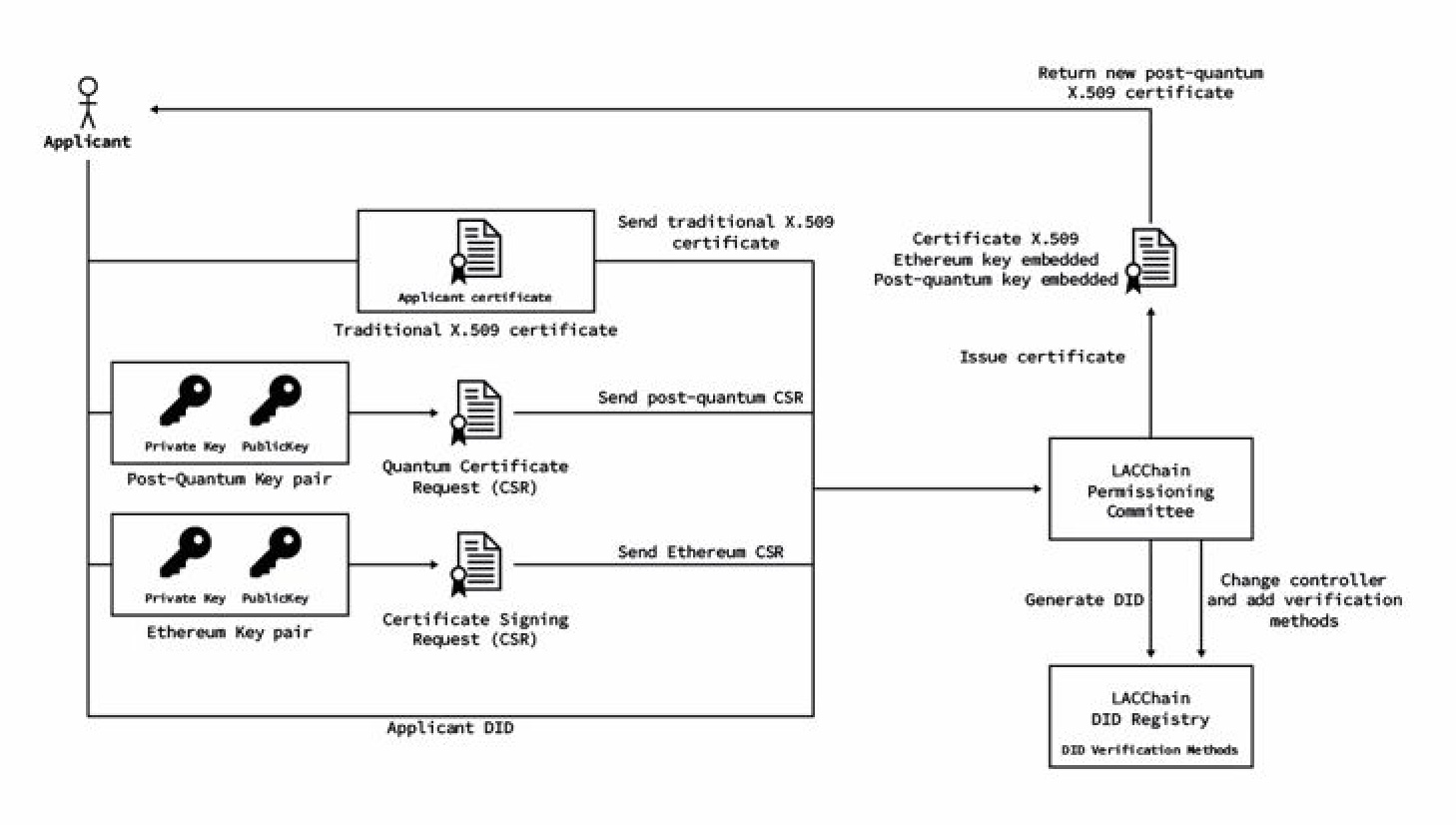}
\caption{{\small High level diagram of the post-quantum certification and on-chain registration of an entity.}}
\label{figure-05}
\end{figure}

The process for the generation of post-quantum certificates is summarized in Figs. (\ref{figure-04},\ref{figure-05}) and broken down into the following seven steps: 

\begin{itemize}

\item 
The applicant requests and receives the entropy form the qRNG as explained in Section \ref{gen-dist-quantum-entropy}.

\item 
The applicant generates a post-quantum Falcon-512 key pair using the quantum entropy through a modified version of the OpenSLL CLI (this modification has been made by the Open Quantum Safe Initiative and we have contributed with a Debian package to simplify its installation) and builds a certificate signing request (CSR).

\item 
The applicant generates a second CSR with an Ethereum key pair that will be used to sign transactions using the default method set by Ethereum (currently ECDSA).

\item 
The applicant sends to a certificate authority (CA) -a role played by the LACChain Technical Team in our pilot- (i) a traditional X.509 issued by a trusted CA, (ii) a certificate signing request (CSR) for the Ethereum key, and (iii) a CSR associated for the Falcon post-quantum key.

\item 
The CA verifies that (i) the traditional X.509 is valid, (ii) the subject in the traditional X.509 matches the subject in the CSRs, and (iii) the signature of the CSRs matches the public keys that are requested to be certified (i.e., the CSRs are valid).

\item 
If the verification fails, the certification process is rejected, and an error message is returned to the applicant.

\item

If the validation process is passed, the CA proceeds to register three items into the smart contract within the blockchain called \lq \lq the Decentralized Identifier (DID) Registry.'' DIDs are URIs that follow a W3C standard \cite{did-w3c}, which are suitable for the identification of individuals, entities, or other components within decentralized environments such as blockchain networks. The three items registered in the smart contract are (i) the DID, (ii) the Ethereum and Falcon  post-quantum  public keys, and (iii) the subject data or alternatively a proof of the subject's identity that does not reveal subject data. Simultaneously, the CA also returns several items to the applicant, including the Falcon post-quantum X.509 certificate that contains the Ethereum public key, the Falcon  post-quantum  public  key, and a new DID controlled by another DID derived from the ETH key.

\end{itemize}

Each of these steps is essential and additional useful clarifications are listed below:

\begin{itemize}

\item

CSR are files of encoded text that contain information to be included in the requested certificate such as the organization name, common name (domain name), address, and country. It also contains the public key that will be included in the certificate, but the private key is not disclosed. Instead, the private key is used to sign the request so the CA can verify that the requester is indeed in control of that particular private key.

\item
The applicant is required to present a traditional X.509 so the blockchain CA does not have to accomplish the verification of the applicant's identity from scratch. Both the applicant and the CA take advantage of a previous X.509 and the CA only verifies that the certified subject data in the X.509 matches the subject data in the CSRs.

\item
The DID Registry follows the DID standard from the W3C \cite{did-w3c} which presents a data model for identifiers particularly designed to be resolved and verified in decentralized registries. Every time the CA certifies a new entity, it registers the DID in the blockchain with the information about the certified Ethereum and Falcon public keys, so that anyone with access to  the public blockchain ledger can resolve the entity's DID and verify the keys associated with them. For example, this would occur when the entity is using the Ethereum key, the Falcon key, or both to  sign a transaction,  which will be addressed  in Subsection \ref{onchainverification}.

\end{itemize}

\subsection{Encapsulation of the communication between nodes using quantum-safe cryptography}
\label{encapcommquantumsafe}

Communication between nodes is made through the protocol established by the blockchain technology and varies depending on the network used. In the case of the LACChain Besu Network used for this pilot, nodes communicate via TCP and use the RLPx for data encryption (this is the same for the Ethereum mainnet, as Hyperledger Besu is an Ethereum client.) This protocol seals messages with a ECDSA signature on curve SECP251k1 to link the network message to a peer address. We decided not to modify this protocol because that would require maintenance of a new blockchain technology. Instead, our goal was to keep using the Hyperledger Besu technology and develop a layer on top to make it quantum-resistant.

With the aim of developing a layer-2 solution that could be used by any blockchain with any communication protocol and that would not be invasive to the protocol (i.e., does not require layer-1 modifications),  our solution consist   of adding a point-to-point TLS tunnel modified to support post-quantum keys where the post-quantum X.509 certificates described in Subsection \ref{generationofpqcs} are used for identification and authorization.

For the pilot, we  used Falcon-512 asymmetric keys.  As this is built on a TLS connection that is not sensitive to the key length, unlike blockchain transactions, it is possible to use other post-quantum algorithms. However, in order to be consistent with the use of a single post-quantum algorithm in the different phases of the pilot, we used Falcon-512.

Once this tunnel is established,  each node must route the traffic aimed at its counterpart through the TLS tunnel, making it unfeasible for a quantum computer to intercept the traffic and impersonate a node. This protects the blockchain network from different types  of  attacks.  For  example,  because  we  are not modifying the blockchain protocol in our permissionless network, the node producers that vote for the generation of new blocks are still materializing this vote in an ECDSA signature (the consensus protocol requires 2/3+1 of node producer's signatures for a block to be considered valid) that is neither replaced not complemented with a post-quantum signature.   However,  if a hacker was to discover all the private ECDSA keys of the validator nodes and tried to tamper with the block production by changing the valid transactions and use the validator nodes' signatures to sign them, it could not achieve it because it cannot intercept the communication between nodes where they could provoke this type of man-in-the-middle attack. The hacker would need to hack and access each of  the validator node servers, for which quantum computers present no advantage.

In any case, despite the fact that we believe this threat is removed with our solution, it would be easier and more convenient to modify the Ethereum protocol so cryptographic algorithms different from ECDSA, such as Falcon-512, are recognized and can be used by validator nodes to sign blocks.

\subsection{Signature of transactions using post-quantum keys}
\label{signature-transactions-pqk}

Unlike the first three phases, the implementation of the fourth phase requires us to be particular about each specific blockchain network. There are blockchain protocols that recognize different encryption algorithms and/or are already flexible in incorporating new ones. At the present moment, this is not the case of Ethereum and the Ethereum-client, Hyperledger Besu, on top of which the LACChain Network used in the pilot is built \cite{besu}. In this context, our way for introducing a mechanism to add a quantum signature to the transactions broadcasted to the network without modifying the blockchain protocol was the development of a relay signer and a meta-transaction signing schema.

A meta-transaction is a mechanism through which to wrap a regular transaction into another transaction addressed to a method of a smart contract (a.k.a. relay Hub) which unwraps and executes the original transaction. Because the meta-transaction is a regular call to a smart contract, we can add new parameters along with the original transaction. In this case, our design allows us  to add the writer node's URI (a DID \cite{did-w3c}) and a post-quantum signature to the original transaction. 

\begin{figure}[hbt]
\includegraphics[width=1\textwidth,center]{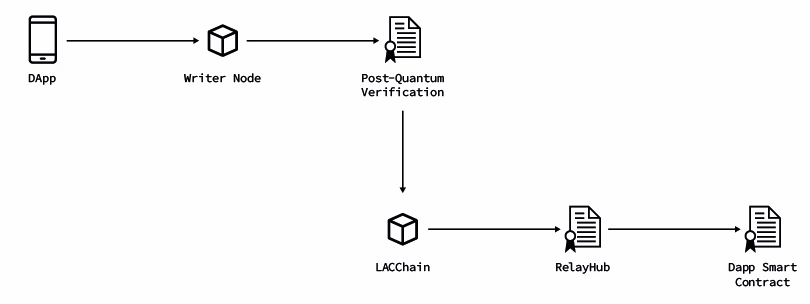}
\caption{{\small High level diagram presenting the different components from the DApp (it can also be an app or any application connected to the writer node and generating transactions) and the smart contract that it is calling.}}
\label{figure-06}
\end{figure}

\begin{figure}[hbt]
\includegraphics[height=0.4\textwidth,center]{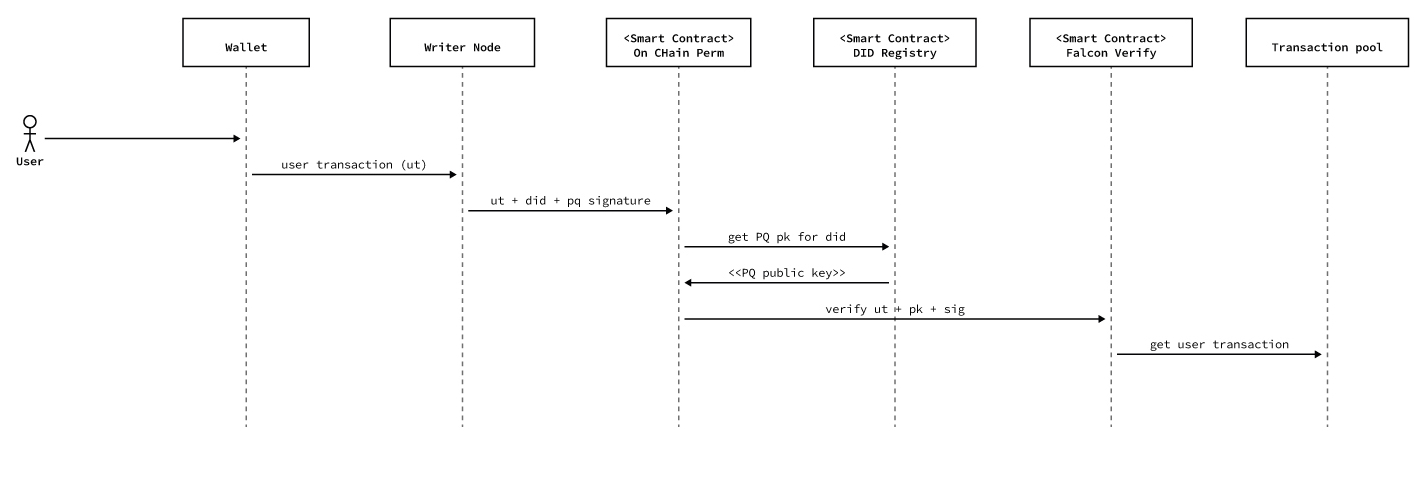}
\caption{{\small High level diagram illustrating the flows from the generation of a transaction to the incorporation of that transaction to the transaction pool of a node, after validating the post-quantum signature.}}
\label{figure-07}
\end{figure}

We have developed a relay signer that is provided to the writer nodes -the only nodes allowed to broadcast transactions according to the LACChain topology \cite{lacchain-topology}- that can manage post-quantum keys. This component exposes a JSON-RPC standard interface, instrumenting methods to make the whole operation transparent to the user. Each writer node is responsible for keeping its Falcon-512 private key safe, and the signer to generate the meta-transaction. Figure \ref{figure-06} summarizes these concepts. Furthermore, full interaction among components is presented in Fig. \ref{figure-07}.

Following the EIP-155 \cite{ethereum-eip-155}, signatures in Ethereum take nine RLP encoded elements: nonce, gasprice, startgas, to, value, data, chainid, 0, 0. For consistency, we took the same stream of data to generate the Falcon-512 signatures. This guarantees the integrity of the original transaction -the writer node cannot modify it- and its quantum resistance by adding the post-quantum signature in the meta transaction. Writer nodes leverage the post-quantum public keys certified by a CA in the post-quantum X.509. 

It is worth mentioning that we are only adding a post-quantum signature in the meta transaction that is created by the writer node, but original senders (i.e., blockchain addresses) are still using only the ECDSA signatures to sign their transactions. Ethereum addresses are the 20 bytes of the SHA3 hashed ECDSA public key, so the public key is not directly exposed. However, when an address sends a transaction, the private key is used to sign it and therefore it is necessary to reveal the public key so the transaction can be verified. 

Thus, if a blockchain address is in possession of certain tokens or has a particularly relevant role in the network (e.g., being permissioned in a smart contact that can issue digital bonds), a quantum computer could be used to hack the private key associated to that address and send transactions to the blockchain that impersonate the true owner. This would allow the hacker to steal the victim's funds or to assume their particularly relevant role in the network, respectively.

Our solution allows to remove this threat by enabling each smart contract to require post-quantum authentication and leveraging for it one of our on-chain verification mechanisms presented in Section \ref{onchainverification} . Only the transference of Ether would not be protected, but LACChain does not have Ether enabled.

As in the case of the signatures by validator nodes described in Section \ref{encapcommquantumsafe}, it would be much easier, ideal, and convenient to have the Ethereum technology enabling the use of quantum-safe cryptographic algorithms that can be used at the protocol level to sign and verify transactions. We believe that Ethereum Improvement Proposals (EIPs) such as the EIP-2938 \cite{ethereum-eip-2938} are moving in the right direction and are very aligned with the work described in this paper.


\newpage{}

\subsection{On-chain verification of post-quantum signatures}
\label{onchainverification}

When a writer node adds a post quantum signature to the meta-transaction and broadcasts it to the network, there must be a mechanism for the signature to be verified. In the regular Ethereum protocol, there is not explicit verification for any signature. In the Ethereum protocol, for a given ECDSA signature, an address is derived and used as the identity of the person willing to execute and pay for a blockchain operation. For the LACChain Besu Network, we have decided to implement a verification protocol based on the Onchain Permissioning feature, which is based on smart contracts. This feature enables each node to intercept every transaction and run different validations before incorporating them into their transaction pool and replicate them to their peers. 

Particularly, according to our protocol, nodes use the post quantum signature to verify the authenticity and integrity of the transaction. As the name of the feature implies, this is resolved by making a local call to a smart contract existing in the network, which receives  several parameters (sender address, target address, transaction value, gas price, gas limit, payload). To our purpose, nodes check the \lq \lq target address'' and dissect the \lq \lq payload'', as described below.

As previously discussed  (see Section \ref{signature-transactions-pqk}), we use a meta-transaction model for executing user requests. This means that there is a single-entry point for our network, which is the address of the Relay Hub contract where the meta-transaction is directed. Therefore, the first Permissioning check consists of verifying that the target address is the Relay Hub contract. Otherwise, nodes will reject the transaction.

Once the Relay Hub smart contract has been verified as the target of the transaction, each node extracts the original payload transaction, the writer node's DID, and the Falcon-512 signature from the original transaction in order to verify the signature. Additionally, a call to the DID Registry allows for retrieval of the public keys associated with it, including the post-quantum public key that should match the post-quantum signature. With this information, each node receiving a transaction from a peer takes the original transaction, the public key, and the signature, and verifies their consistency. If it is not consistent, they reject the transaction (i.e., they do not add it to their transaction pool, nor propagate it to other peers).

To summarize, the protocol we have designed consists of three steps:

\begin{enumerate}

\item 
Every node that receives a meta-transaction -from the node that created it or from another node that replicated it- checks the sender. This involves obtaining the DID from the meta-transaction and locally querying the DID Registry in order to resolve (i.e., obtain) its Ethereum keys (ECDSA). They then verify that the public key derived from the ECDSA signature of the meta-transaction has control over the  node's DID that generated it.

\item 
If Step 1 is successful, the node calls the DID Registry again and now resolves the post-quantum public key associated with the DID as well as the Ethereum public key verified in Step 1.

\item 
With the post-quantum public key resolved from the DID Registry in Step 2, the post-quantum signature, and the original transaction, each node then verifies the post-quantum algorithm.

\end{enumerate}

If the three previous steps are successfully completed, nodes add the meta-transaction to their transaction pool and replicate them onto other nodes so that the validators will receive them and add them into the next block.

As previously stated, we have chosen Falcon-512 as our post-quantum algorithm.  There is not yet an ideal way of implementing the Falcon-512 verification required to accomplish the Step 3 of this verification process nor any other post-quantum algorithm, in Ethereum-based networks. We have developed three alternative mechanisms and analyzed their pros and cons,  which are presented in detail in Subsection \ref{comparisonpostquantumsignatures}. 

These three mechanisms are:
\begin{itemize}

\item 
Implementing the verification code in Solidity (see Subsection \ref{verification-code}).

\item 
Implementing solidity instruction in the Solc compiler and corresponding EVM opcode, written in Java (Besu is written in Java), that performs a call through JNI to a NIST-compliant and high performance native Liboqs library outside of the EVM virtualized environment (see Subsection \ref{evm-virtual-machine}).

\item 
Refactoring the EVM opcode Java from the EVM virtual machine into a pre-compiled contract (a EVM Java-code native smart contract) that performs the call through JNI to the NIST compliant,  high performance native Liboqs library outside of the EVM virtualized environment (see Subsection \ref{evm-pre-compiled}).

\end{itemize}
We hope that in the not-so-distant future, we can use this effort in alignment with the upcoming protocol changes in the form of the Accounts Abstractions, which will allow us to replace ECC cryptography with new algorithms, including post-quantum.

\subsubsection{Verification code in solidity} 
\label{verification-code}

The natural execution environment for the blockchain is the Ethereum Virtual Machine; thus, in our first attempt, we implemented the verification code entirely in the Solidity language. We dissect the reference implementation in the following modules and discuss the implementation of the highlighted functions one by one.

\begin{figure}[hbt]
\includegraphics[height=0.75\textwidth,center]{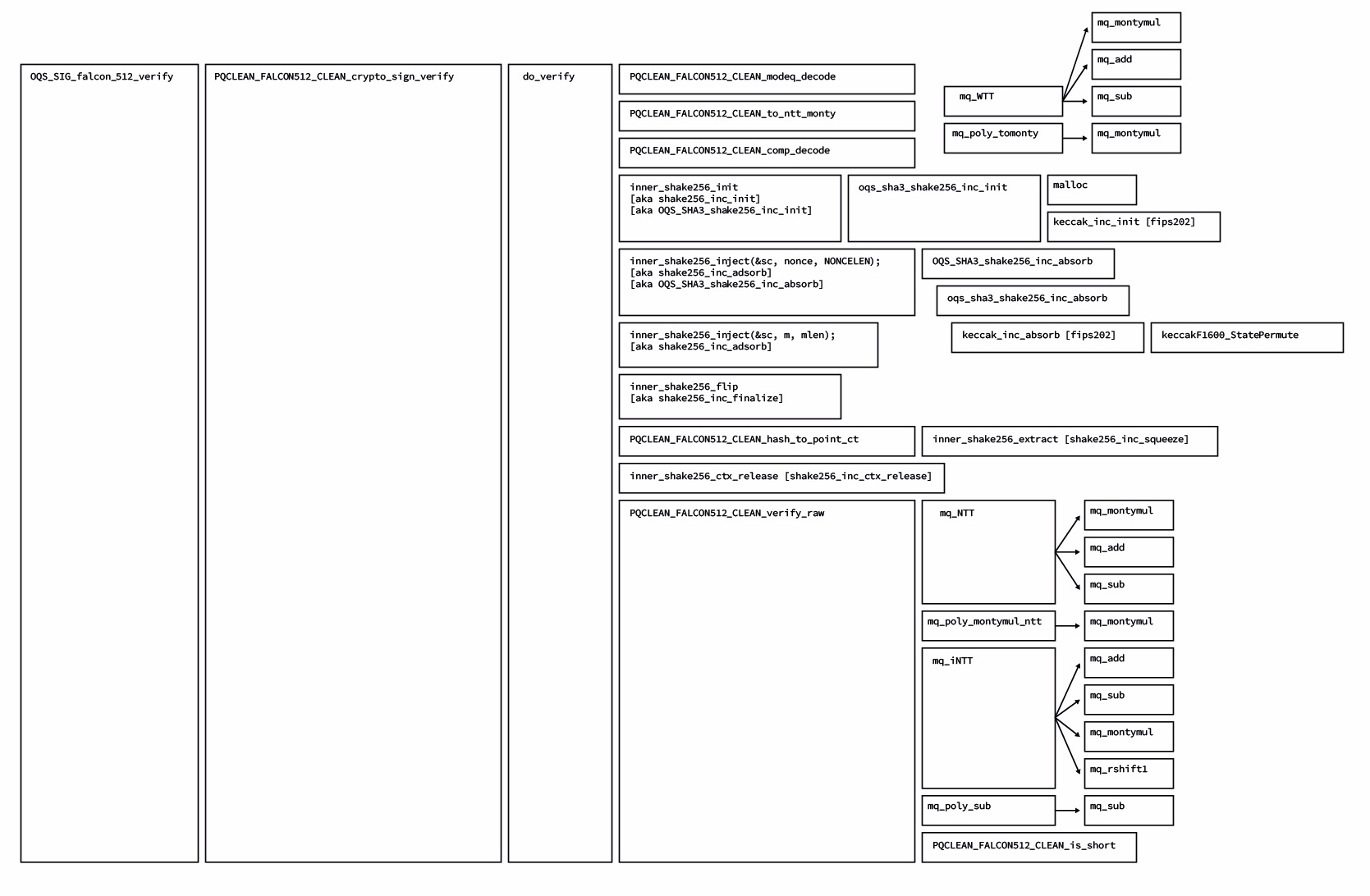}
\caption{{\small High level function hierarchy of Falcon highlighting the necessary calls for verification.}}
\label{figure-08}
\end{figure}

\begin{figure}[hbt]
\includegraphics[width=1\textwidth,center]{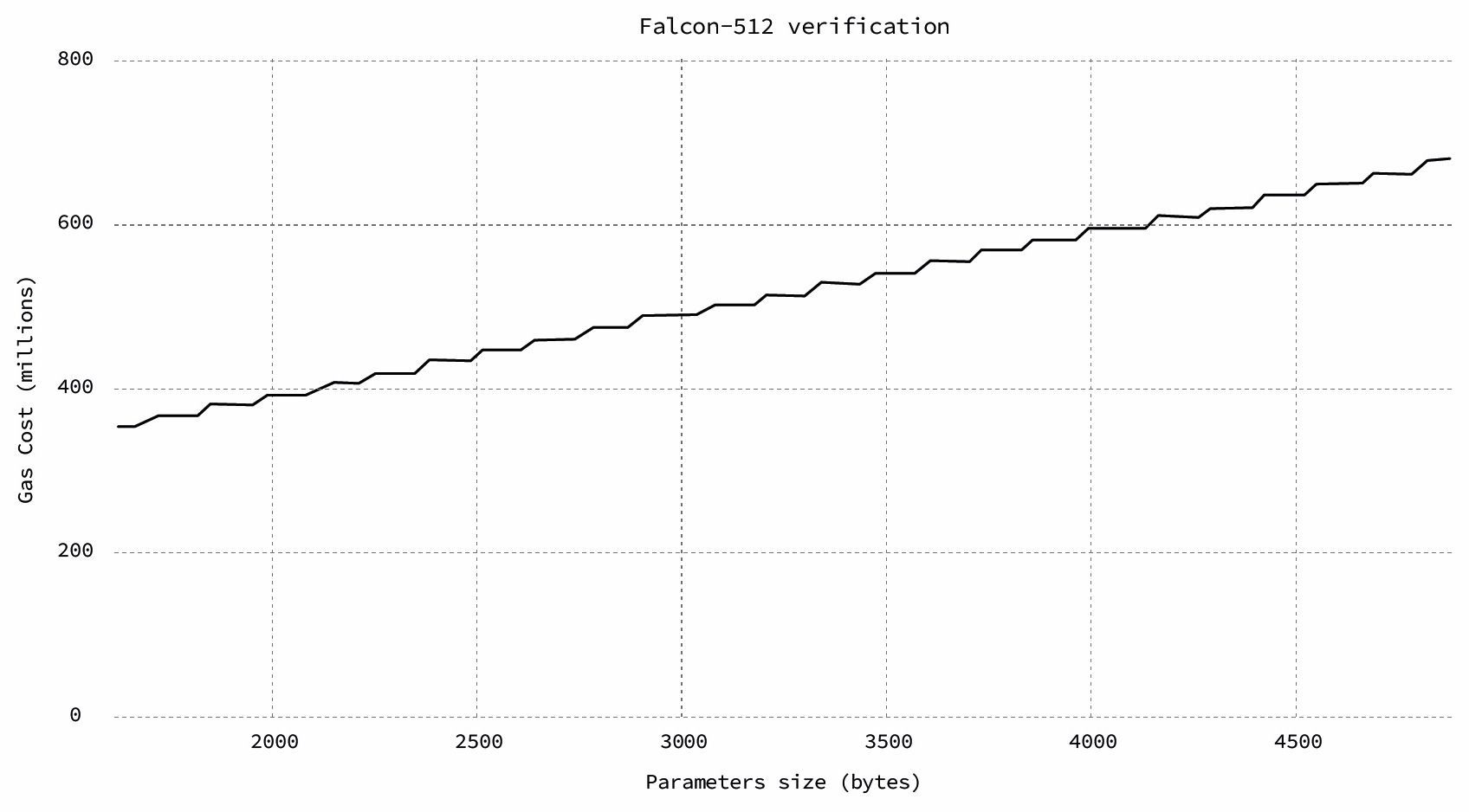}
\caption{{\small Gas consumption by the on-chain verification of Falcon-512.}}
\label{figure-09}
\end{figure}

Implementing the highlighted portions of Fig. \ref{figure-08}  in Solidity allowed for on-chain signature verification. Upon the completion of the development process, we faced two major problems. The first problem was the code size. It exceeded the 24kb limit that Ethereum mainnet imposes. This limit could have been exceeded in LACChain because LACChain has different boundaries, but such large code sizes are not ideal. The second and more major problem was the execution cost. In  Fig. \ref{figure-09}, we present a chart with the execution cost of the verification of the known answer tests provided by the Falcon implementation. If we compare the average 500 million gas units for a single Falcon signature verification, with the current block limit of 12 million gas units in the Ethereum mainnet, we can conclude that this approach is completely impractical at this point.

\subsubsection{EVM virtual machine-based signature validation support} 
\label{evm-virtual-machine}

An EVM based approach requires modification of both the Solidity compiler (solc) and the Ethereum Virtual Machine (EVM) that underpins the Besu Hyperledger technology used by LACChain.

These changes are applicable across all Ethereum-based networks but require all participating nodes within the blockchain to utilize the updated solidity compiler and EVM. The Java Native Interface (JNI) is also required in addition to ensuring that compatible OpenQuantum Safe (an open-source venture) Liboqs libraries are installed. Performance is therefore limited only by the native liboqs library and the native node processing power.

The solidity modification is minor, and only requires adding an instruction token to the existing instruction list. The modification to the EVM is similarly minor and only requires adding a Java class to a Falcon Verify operation and registering the class with the operations available for that version of the EVM virtual machine. This implementation provides a simple Gas cost of 1. However, an extended example could be made to utilize the memory-block size cost calculation performed by SHA3.

\begin{figure}[hbt]
\includegraphics[height=0.4\textwidth,center]{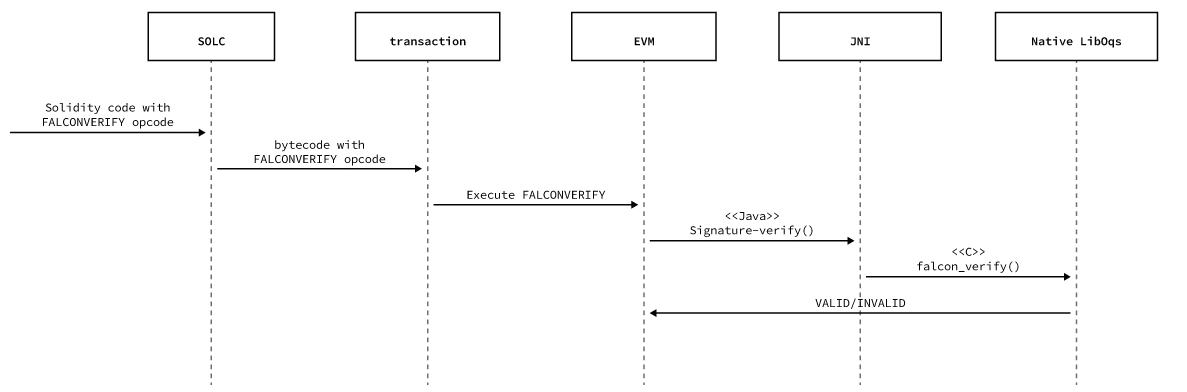}
\caption{{\small EVM virtual machine-based signature validation support.}}
\label{figure10}
\end{figure}

\begin{figure}[hbt]
\begin{center}
\includegraphics[width=1\textwidth,center]{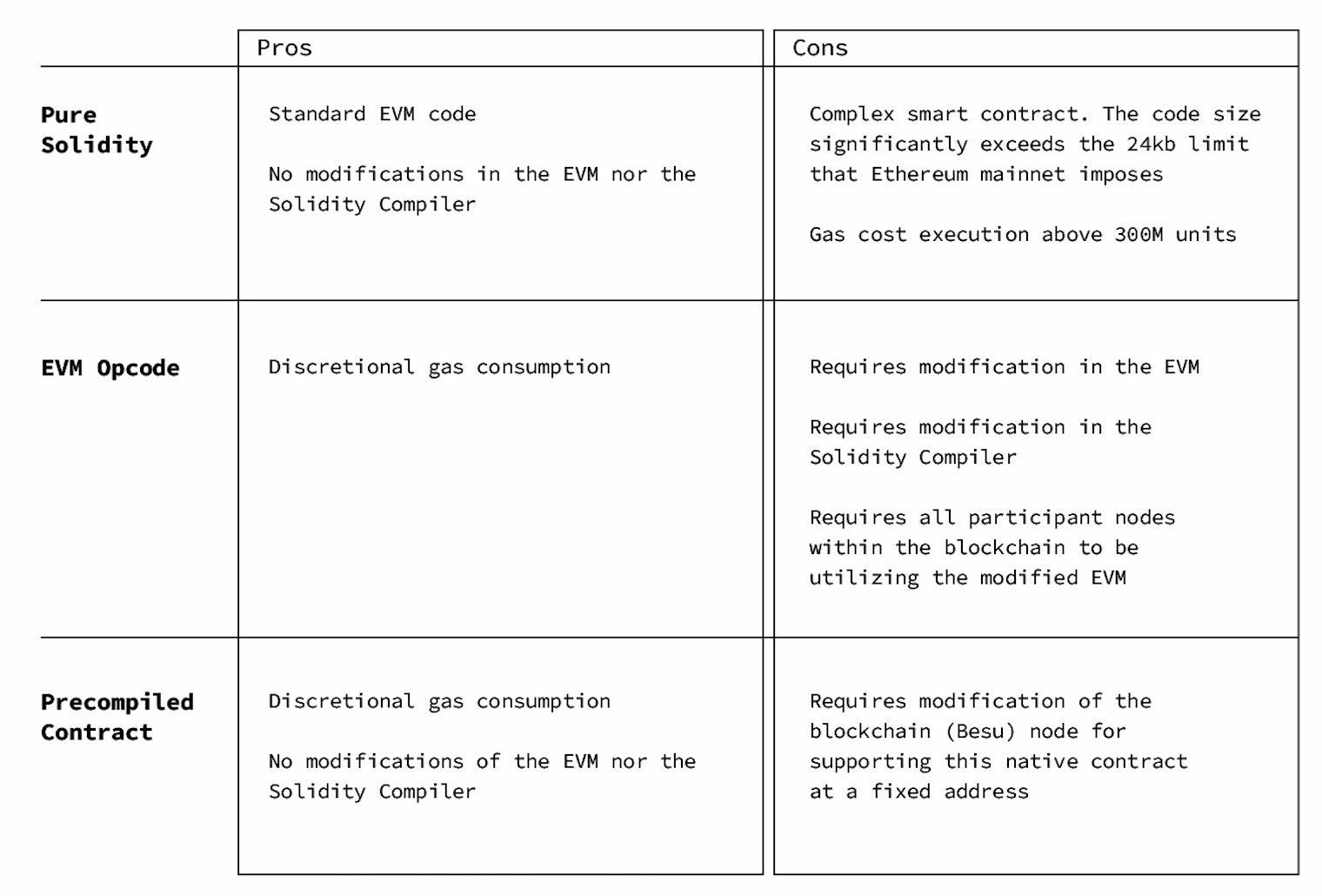}
\end{center}
\caption{{\small Pros and Cons of Pure Solidity, EVM Opcode, and Precompiled contract.}}
\label{figure-11}
\end{figure}

The approach only uses one opcode from the 6000 opcodes limit call within the standard configuration of Ethereum. The real-world performance of the signature verification is as fast as the hardware can perform - aligning with the performance observed by the OpenQuantum Safe teams.

The utilization of the OpenQuantum Safe liboqs library ensures minimal operational delay or risk in maintaining updated quantum algorithms in line with NIST and the OpenSource Safe current standards. The Java class implemented for the EVM can also be extended beyond Falcon-512 and to allow Falcon-1024 or other signatures.

The EVM stack word width is 256bits, which naturally fits with the existing 256-bit hashes used in the classical encryption. However, post-quantum signatures with larger memory requirements will become less optimal unless the stack word width is increased at the cost of compatibility with previously operational blockchains. Finally, the POC EVM implementation utilizes Falcon-512, which minimizes this impact while also providing a security level that is in alignment with classical AES-256. Fig. \ref{figure10} summarizes the interactions described in this subsection.

\subsubsection{EVM pre-compiled-based signature validation support} 
\label{evm-pre-compiled}

The pre-compiled approach transplants the EVM falcon verify operation Java class into a EVM precompiled smart contract (a native Java compiled smart contract). This approach has two benefits that reduce operational impact:

\begin{itemize}
\item
No change to the Solidity compiler.

\item
No change to the underlying EVM virtual machine.
\end{itemize}

This facilitates the distribution of the quantum signature verification separate from the compiler and EVM releases. The approach therefore brings all the benefits of the EVM opcode implementation  but with less operational work. The JNI and Liboqs libraries are used identically, offering speed and ease of maintenance. It is also worth mentioning that given this verification is meant to be executed before a node joins the blockchain, it could easily be replaced in the future without affecting the consensus. It will only be necessary to modify the deployment scripts. 

Implementing this solution in the LACChain Hyperledger Besu Network would require changes in the protocol with respect to other Ethereum networks, including the mainnet. This would be against our goal to preserve compatibility with the Ethereum community. Therefore, the ideal way to proceed with this third approach for the verification of Falcon signatures is submitting an EIP for the community to evaluate the incorporation of a pre-compiled smart contract into the Ethereum protocol, being this either the full Falcon verification algorithm or same detected bottlenecks from a gas consumption perspective.

Fig. \ref{figure-11} shows some advantages and disadvantages of Pure Solidity, EVM Opcode and precompiled contract.

\subsubsection{Comparison between different solutions for verification of post-quantum signatures} \label{comparisonpostquantumsignatures}

The three alternatives that were designed and tested for the verification of post-quantum signatures are successful for verification but not ideal for a productive implementation if the Ethereum-based network implementing them is intended to remain fully compatible with the Ethereum mainnet. The Solidity native implementation presented in Subsection \ref{verification-code} is not scalable due to the amount of gas required for the execution of the code, although it does not require a modification of Besu or Ethereum. The modification of the Solidity compiler and the EVM, as well as the pre-compiled smart contract (presented in Subsections \ref{evm-virtual-machine} and \ref{evm-pre-compiled} respectively) are computationally scalable.  However, they require undesired modifications unless otherwise agreed upon by the entire Ethereum community, which is the goal we aim at to pursue in the next step of this pilot.

Additionally, the solutions described in Subsections \ref{evm-virtual-machine}  and \ref{evm-pre-compiled} use the Java Virtual machine. However, unlike the Solidity native implementation, these two techniques are not impacted by EVM or JavaVM mathematical computational problems maintaining validity and security between releases. Instead, the pure C native method of Liboqs implements its own mathematical validity tests as part of the C build system. The result is that regardless of Java or EVM release, the verifying Liboqs library remains mathematically valid (assuming no optimizations or changes that invalidate tests). This approach allows organizations to separate security requirements, offering  more precise maintenance and governance. However, this approach would require extra security protocols with the additional overhead.

\section{Conclusions and next steps}
\label{conclusions}
We have analyzed the various areas of blockchain technology threatened by the advent of quantum computers and identified two areas that are under particularly critical risk: internet communication between blockchain nodes and the blockchain transaction signatures. Today, these protocols rely on  algorithms such as ECDH and ECDSA, which are susceptible to attacks  by quantum computers. Current quantum computers have already proven themselves able to break short asymmetric keys using Shor's algorithm and it is only a matter of time before robust quantum computers  currently under development will be able to break larger and larger keys. As the \lq \lq hack today, crack tomorrow'' motto warns, quantum computers will be able to access secrets retroactively. This is particularly critical for blockchain, where information is recorded publicly and immutably so having access to all the information any time in the future will not even require any hacking.

The scarce previous work on this topic has focused on theoretical approaches, with the exception of one implementation of a QKD scheme for key establishment, which requires nodes to run close together on-site due to QKD channels’ length constraints. In this paper, we have proposed the first robust and scalable solution, to our knowledge, to protect communications and signatures in a blockchain network from attacks by quantum computers. Its effectiveness has been demonstrated by its implantation in a real blockchain network. Our solution consists of modifying libSSL to incorporate post-quantum algorithms that are quantum-resistant and adding post-quantum keys into X.509 certificates derived from traditional certificates. The nodes use these post-quantum X.509 certificates to encapsulate their communication by establishing post-quantum TLS tunnels. The nodes also use the post-quantum key associated with the certificate to sign the transactions they broadcast to the network. We have implemented this solution in the LACChain Besu Network, which is built on Ethereum technology.

There are several strengths and benefits to our implementation. First, it uses a quantum source of entropy (i.e., a non-deterministic quantum random number generator) as the seed for the generation of post-quantum keys. Second, we are respectful of the exchange of transactions and the blockchain protocols for encryption discovery and communication (still happens inside the post-quantum tunnel). Third, we have proposed three different alternatives for the post-quantum signature verification, which every node accomplishes before adding a transaction to the transaction pool and replicating it. Therefore, if a signature is not valid, the transaction is never propagated nor added into a block.

The three different solutions for the verification of the post-quantum signatures that we have proposed, developed, and tested are: an implementation of the verification code in Solidity, the addition of a new operation code into the EVM assembly language (with a corresponding Solidity compiler modification to generate this _opcode_), and the introduction of a new pre-compiled (i.e., native) smart contract. The first solution, despite the fact that it is totally compatible with the current protocol, is not computationally scalable due the enormous gas cost it involves. The latter two were implemented through a native Liboqs library outside of the EVM runtime allowing us to improve the execution time and to adjust gas consumption. The experience gathered through this work will lead our team to raise the discussion through an EIP to support the use of Falcon-512 for on-chain verifications. This is the way to not diverge LACChain or any other particular blockchain network from Ethereum consensus and, at the same time, improve the security of any implementation of the protocol.

The three different solutions for the verification of the post-quantum signatures that we have proposed, developed, and tested are: an implementation of the verification code in Solidity, an implementation of the verification in a native Liboqs library outside of the EVM virtualized environment, and a verification using a EVM Java-code pre-compiled (i.e., native) smart contract. These three implementations are focused on ensuring the minimization of the number of operations and amount of entropy required, in addition to being NIST compliant. The first solution is not computationally scalable, and the other two require modification to the Ethereum protocol, which we will propose to the Ethereum community in the form of an EIP. 

In addition to the potential modifications of the Ethereum protocol to enable our layer-two implementation, we also believe it is necessary to modify current blockchain protocols to introduce new post-quantum signature cryptographic algorithms that allow the use of post-quantum cryptography natively. We hope that our work can contribute to current efforts in this direction such as the EIP-2938.

With respect to other blockchain networks that  are  not Ethereum-based, the proposed solution for a quantum-safe blockchain network presented in this paper is applicable too. However, the solution implementation will vary based on the technology used. Therefore, this solution might enable quantum-safeness in other blockchain networks in a more efficient way than in the Ethereum-based network. 

\section*{Acknowledgements}
We gratefully acknowledge the review and comments provided by Ignacio Alamillo, Solomon Cates, Suzana Maranh\~ao Moreno, and Marta Piekarska-Geater. Furthermore, we warmly thank the support of Nuria Simo and Irene Arias. SEVA thanks his family for their unconditional support.

\bibliography{Quantum-BlockChain-Arxiv-June-2021}

\newpage{}
\begin{figure}
 \centering 
\includepdf[pages={1,1}]{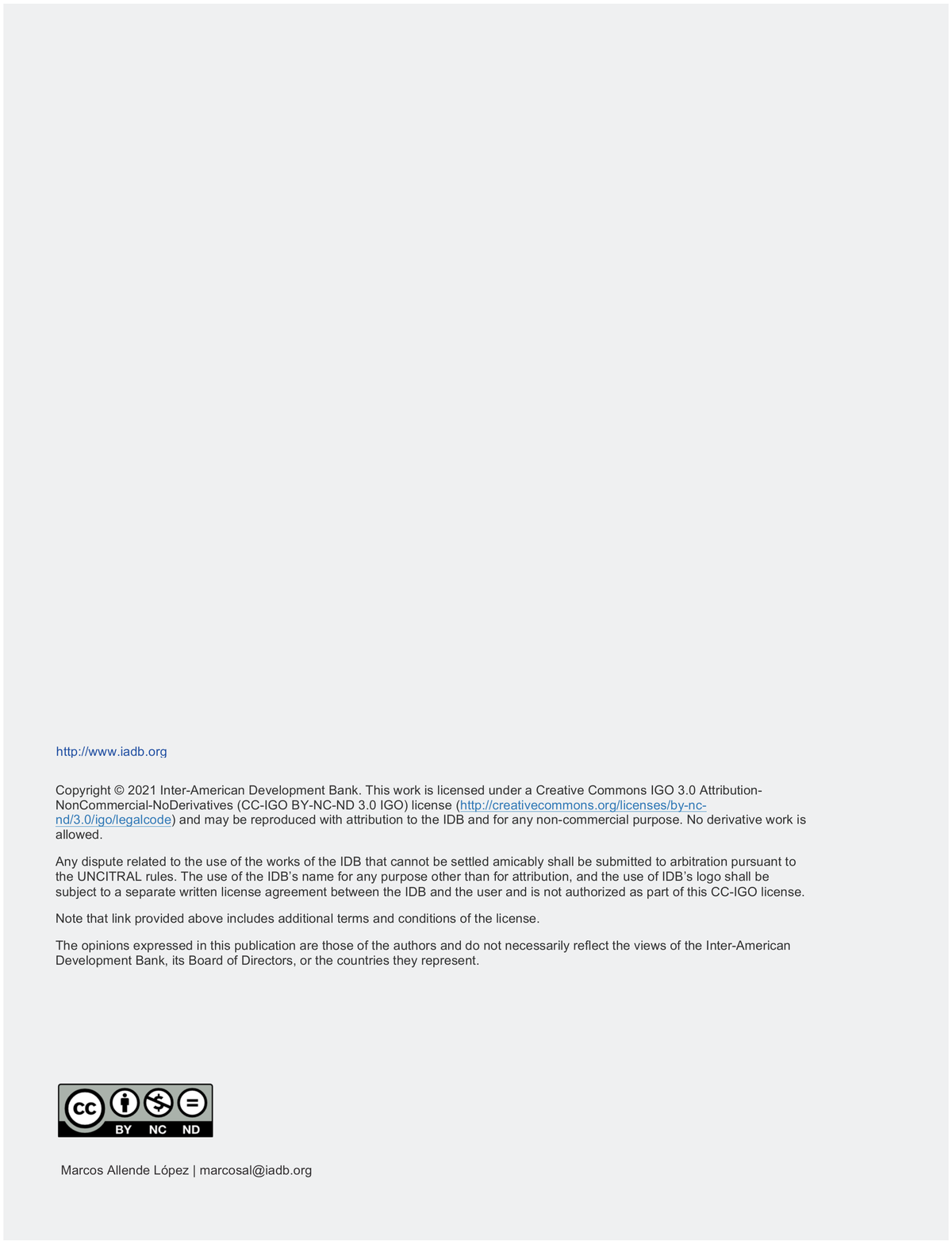}  
\end{figure}
\end{document}